# Estimation of the Location-Scale Parameters Based on Ranked Set Sampling Combined with Additional Binomial Information

Moh'd Alodat[1] and Ayman Baklizi[2]

[1]Department of Statistics, Sultan Qaboos University, Muscat, Oman

[2]Department of Mathematics, Statistics and Physics

Qatar University, Doha, Qatar

Emails: m.alodat@squ.edu.om[1] and a.baklizi@qu.edu.qa[2]

**Abstract:** In this paper, we propose a modified version of ranked set sample which allows for the incorporation of more information in the inference procedure at almost no cost. This procedure is studied for the location-scale family using a likelihood framework. Our results show that the modified procedure produce more accurate inferences than the original ranked set sampling scheme.

## 1. Introduction

The ranked set sampling (RSS) was introduced by McIntyre (1952) as an alternative to the simple random (SRS) in order to increase the efficiency of an experiment. When applying RSS to some population, it is assumed that the sampling units are difficult to measure but can be easily ranked via visual mechanism or any other cheap method. In other words, it assumes that ranking and sampling cost is negligible. This sampling scheme combines visual judgment with actual measurements to increase the precision of the estimation of the population mean. In order to introduce the reader to the McIntyre' RSS, we need to define the order function on a sample. Let $S_m = \{s_1, \ldots, s_m\}$ be a sample of size $m$ from the population of interest. A function $\pi: \{1, 2, 3, \ldots, m\} \to S_m$ is called an order function on $S_m$ if for every $i = 1, 2, \ldots, m$, $\pi$ maps $i$ to the $i^{th}$ largest value in $S_m$, i.e. $s_{\pi(1)} < \cdots < s_{\pi(m)}$.

McIntyre proposes to select an independent but not identically distributed according to the following steps:

Step 1: Randomly select $m^2$ sample units from the population of interest.

Step 2: Allocate the $m^2$ selected units as randomly as possible into $m$ sets(rows) each of size $m$.

$$\begin{matrix} X_{11} & X_{12} & \cdots & X_{1m} \\ X_{21} & X_{22} & \cdots & X_{2m} \\ \vdots & \vdots & \ddots & \vdots \\ X_{m1} & X_{m2} & \cdots & X_{mm} \end{matrix}$$

Step 3: Without yet knowing any values for the variable of interest, rank the units within each set based on a perception of relative values for this variable. This may be based on a visual inspection or done with measurement of a covariant that is correlated with the variable of interest. Ranking each row visually yields.

$$\begin{matrix} X_{1\pi(1)} & X_{1\pi(2)} & \cdots & X_{1\pi(m)} \\ X_{2\pi(1)} & X_{2\pi(2)} & \cdots & X_{2\pi(m)} \\ \vdots & \vdots & \ddots & \vdots \\ X_{m\pi(1)} & X_{m\pi(2)} & \cdots & X_{m\pi(m)} \end{matrix}$$

Notice that $X_{i\pi(j)}$ represents the $j^{th}$ order statistic of the $i^{th}$ set.

Step 4: Select the RSS of size $m$ for actual analysis that consists of the smallest ranked unit from the first set, the second smallest ranked unit from second set, continuing in this fashion until the largest ranked unit is selected from the last set. The diagonal elements $\{X_{1\pi(1)}, X_{2\pi(2)}, \ldots, X_{m\pi(m)}\}$ represents the RSS of size $m$. This process can be repeated $r$ times to get a ranked set sample of size $rm$. Such sample will be denoted by $\{X_{i\pi(i)}^{(j)} | i = 1, 2, \ldots, m; j = 1, 2, \ldots, r\}$ The advantage of this sampling scheme is that it allows the sampler to use his visual judgment to get more information at almost no cost, resulting in a more precise inference.

The McIntyre's RSS scheme has been modified by several authors in order to increase the efficiency of estimation when the underlying distribution is parametric. A good survey about RSS can be found in Chen et al. (2004). For example, Samawi et al. (1996) have introduced the notion of extreme RSS to estimate the mean of a symmetric population. As a result they have shown that sample mean of an extreme RSS is more accurate than its counterpart under McIntyre's RSS as estimators for the population mean. Alodat and Al-Saleh (2001) introduced the notion of moving extreme ranked set sampling. They have derived an unbiased estimator for the population mean of symmetric population. Koyuncu (2017) has considered new ratio estimators based on the neoteric ranked sampling scheme. A recent review about other modifications and applications can be found in Al-Omari and Bouza (2014).

The RSS usually assumes that the sampling and ranking costs of population the units are negligible compared with the cost for measuring the variable of interest (Wang et al., 2004). In general, it is known that the relative precision increases as a function in the set size $m$, keeping other factors fixed. So, if ranking is free, we may choose a set size as large as possible such that it does not deteriorate the accuracy or ranking. However, in practice where RSS might be useful, we cannot ignore the cost of sampling and ranking because cost-effectiveness of RSS arises when sampling and ranking many more units than those measured. Therefore, if cost of sampling and ranking cannot be ignored, more cost will be incurred when using only a large sample obtained through the visual grouping. In this paper we modify on the McIntyre's RSS scheme in order to increase the accuracy of inference without increasing the set size $m$. So, we propose the following modification on McIntyre's RSS scheme:

(i). Select a RSS of size $m$, say $X_{1\pi(1)}, X_{2\pi(2)}, \dots, X_{m\pi(m)}$, from the population of interest

(ii). For each $i = 1, 2, \dots, m$, select(independently) a SRS of $m$ units from the population of interest. Then use a visual mechanism to count the number of units among the $m$ units whose values are less than or equal to $X_{i\pi(i)}$. Denote this number by $Z_i$.

(iii). The steps (1) and (2) produce the sample $(X_{1\pi(1)}, Z_1), (X_{2\pi(2)}, Z_2), \dots, (X_{m\pi(m)}, Z_m)$.

(iv). If the steps (1) and (2) are repeated $r$ times, then a sample of size $n = mr$ is obtained. Let us denote this sample by $\mathcal{D} = \left\{ \left( X_{i\pi(i)}^{(j)}, Z_i^{(j)} \right) \mid i = 1, 2, \dots, m; j = 1, 2, \dots, r \right\}$.

It is clear that the additional information contributed by $Z_i^{(j)}$ was obtained at practically no cost. However its incorporation in the estimation process can result in a more precise inference. In this paper we shall investigate this modified procedure for inference about the parameters of the location-scale family of distributions using likelihood methods.

The rest of the paper is organized as follows. The likelihood function and the associated quantities are derived in section 2. An example using simulated data is presented in section 3. A simulation study to investigate and compare the proposed method with the traditional ranked set sample is described in section 4. The results and conclusions are given in the final section.

## 2. Likelihood Inference

Assume that the population has an absolutely continuous distribution function $F(x;\boldsymbol{\theta})$, $\boldsymbol{\theta} \in \mathbb{R}^p$, where $\mathbb{R}^p$ is the p-dimensional Euclidean space. In the steps (i)-(iv), the random variables $X_{i\pi(i)}^{(j)}$ and $Z_i^{(j)}$ are assumed independent. So conditional on $X_{i\pi(i)}^{(j)}$, the distribution of $Z_i^{(j)}$ is binomial, i.e., $Z_i^{(j)}|X_{i\pi(i)}^{(j)} \sim Bin\left(m, F(X_{i\pi(i)}^{(j)};\boldsymbol{\theta})\right)$. It can be noticed that the distribution of $Z_i^{(j)}$ does not depend on $\boldsymbol{\theta}$, i.e., $Z_i^{(j)}$ is ancillary for $\boldsymbol{\theta}$. However, $Z_i^{(j)}$ is correlated with $X_{i\pi(i)}^{(j)}$. So the data $\mathcal{D}$ can be used to improve the inference about $\boldsymbol{\theta}$ based on $X_{i\pi(i)}^{(j)}, i = 1, 2, \ldots, m$. It can be noticed that marginal distribution of $Z_i^{(j)}$ is free of the $F(X_{i\pi(i)}^{(j)};\boldsymbol{\theta})$ and hence is free from $\boldsymbol{\theta}$. So $Z_i^{(j)}$ is ancillary for $\boldsymbol{\theta}$. Although $Z_i^{(j)}$'s are ancillary for $\boldsymbol{\theta}$, they can be used to improve the inference about $\boldsymbol{\theta}$ when they are combined with the $X_{i\pi(i)}^{(j)}$'s.

Let $\mathcal{D} = \left\{\left(X_{i\pi(i)}^{(j)}, Z_i^{(j)}\right) \mid i = 1, 2, \ldots, m; j = 1, 2, \ldots, r\right\}$ be selected according to the sampling scheme proposed by (i)-(iv). Then the likelihood of $\mathcal{D}$ is

$$\mathcal{L}(\boldsymbol{\theta};\mathcal{D}) = \prod_{i=1}^{m}\prod_{j=1}^{r} \binom{m}{Z_i^{(j)}} F\left(x_{i\pi(i)}^{(j)};\boldsymbol{\theta}\right)^{Z_i^{(j)}} \left(1 - F\left(x_{i\pi(i)}^{(j)};\boldsymbol{\theta}\right)\right)^{m-Z_i^{(j)}} \times$$

$$\prod_{i=1}^{m}\prod_{j=1}^{r} i\binom{m}{i} F\left(x_{i\pi(i)}^{(j)};\boldsymbol{\theta}\right)^{i-1} \left(1 - F\left(x_{i\pi(i)}^{(j)};\boldsymbol{\theta}\right)\right)^{m-i} f\left(x_{i\pi(i)}^{(j)};\boldsymbol{\theta}\right),$$

$$= \prod_{i=1}^{m}\prod_{j=1}^{r} i\binom{m}{Z_i^{(j)}}\binom{m}{i} F\left(x_{i\pi(i)}^{(j)};\boldsymbol{\theta}\right)^{Z_i^{(j)}+i-1} \left(1 - F\left(x_{i\pi(i)}^{(j)};\boldsymbol{\theta}\right)\right)^{2m-Z_i^{(j)}-i} f\left(x_{i\pi(i)}^{(j)};\boldsymbol{\theta}\right),$$

where $f(.;\boldsymbol{\theta})$ is the pdf of $F(.;\boldsymbol{\theta})$. Now, we assume that $F(.;.)$ belongs to the location scale family and is twice differentiable with respect to $x$. If $F(x;\boldsymbol{\theta}) = F\left(\frac{x-\mu}{\sigma}\right)$, where $\boldsymbol{\theta} = (\mu,\sigma)^T$, then the log-likelihood is

$$\mathcal{L}^*(\boldsymbol{\theta};\mathcal{D}) = \log \mathcal{L}(\boldsymbol{\theta};\mathcal{D}) = c + \sum_{i=1}^{m}\sum_{j=1}^{r}(z_i^{(j)}+i-1)\log\left(F\left(\frac{x_{i\pi(i)}^{(j)}-\mu}{\sigma}\right)\right) +$$

$$\sum_{i=1}^{m}\sum_{j=1}^{r}(2m-z_i^{(j)}-i)\log\left(1-F\left(\frac{x_{i\pi(i)}^{(j)}-\mu}{\sigma}\right)\right) + \sum_{i=1}^{m}\sum_{j=1}^{r}\log f\left(\frac{x_{i\pi(i)}^{(j)}-\mu}{\sigma}\right)$$
$$- mr\log\sigma.$$

In order to economize in the notation, we introduce the variables $v_{ij} = \left(x_{i\pi(i)}^{(j)}-\mu\right)/\sigma$, $i = 1,2,\ldots,m; j = 1,2,\ldots,r$. If so, then $\frac{dv_{ij}}{d\mu} = -\frac{1}{\sigma}$ and $\frac{dv_{ij}}{d\sigma} = -\frac{v_{ij}}{\sigma}$. Hence the first derivatives of $\mathcal{L}^*(\boldsymbol{\theta};\mathcal{D})$ with respect to $\mu$ and $\sigma$, respectively, are given by

$$\frac{d\mathcal{L}^*(\boldsymbol{\theta};\mathcal{D})}{d\mu} = -\frac{1}{\sigma}\sum_{i=1}^{m}\sum_{j=1}^{r}(z_i^{(j)}+i-1)\frac{f\left(\frac{x_{i\pi(i)}^{(j)}-\mu}{\sigma}\right)}{F\left(\frac{x_{i\pi(i)}^{(j)}-\mu}{\sigma}\right)} +$$

$$\frac{1}{\sigma}\sum_{i=1}^{m}\sum_{j=1}^{r}(2m-z_i^{(j)}-i)\frac{f\left(\frac{x_{i\pi(i)}^{(j)}-\mu}{\sigma}\right)}{1-F\left(\frac{x_{i\pi(i)}^{(j)}-\mu}{\sigma}\right)} - \frac{1}{\sigma}\sum_{i=1}^{m}\sum_{j=1}^{r}\frac{f'\left(\frac{x_{i\pi(i)}^{(j)}-\mu}{\sigma}\right)}{f\left(\frac{x_{i\pi(i)}^{(j)}-\mu}{\sigma}\right)}.$$

$$= -\frac{1}{\sigma}\sum_{i=1}^{m}\sum_{j=1}^{r}(z_i^{(j)}+i-1)\frac{f(v_{ij})}{F(v_{ij})} + \frac{1}{\sigma}\sum_{i=1}^{m}\sum_{j=1}^{r}(2m-z_i^{(j)}-i)\frac{f(v_{ij})}{1-F(v_{ij})}$$

$$-\frac{1}{\sigma}\sum_{i=1}^{m}\sum_{j=1}^{r}\frac{f'(v_{ij})}{f(v_{ij})}$$

and

$$\frac{d\mathcal{L}^*(\boldsymbol{\theta};\mathcal{D})}{d\sigma} = -\frac{1}{\sigma}\sum_{i=1}^{m}\sum_{j=1}^{r}(z_i^{(j)}+i-1)\frac{v_{ij}f(v_{ij})}{F(v_{ij})} +$$

$$\frac{1}{\sigma}\sum_{i=1}^{m}\sum_{j=1}^{r}(2m-z_i^{(j)}-i)\frac{v_{ij}f(v_{ij})}{1-F(v_{ij})} - \frac{1}{\sigma}\sum_{i=1}^{m}\sum_{j=1}^{r}\frac{v_{ij}f'(v_{ij})}{f(v_{ij})} - \frac{mr}{\sigma}.$$

The MLEs for $\mu$ and $\sigma$ are obtained by solving the likelihood equations $\frac{d\mathcal{L}^*(\theta;\mathcal{D})}{d\mu} = 0$ and $\frac{d\mathcal{L}^*(\theta;\mathcal{D})}{d\sigma} = 0$ simultaneously for $\mu$ and $\sigma$ so that the hessian matrix at the solution is negative. It can be noticed that for normal population that the last two equations reduce to:

$$-\sum_{i=1}^{m}\sum_{j=1}^{r}(z_i^{(j)} + i - 1)\frac{\varphi(v_{ij})}{\Phi(v_{ij})} + \sum_{i=1}^{m}\sum_{j=1}^{r}(2m - z_i^{(j)} - i)\frac{\varphi(v_{ij})}{\Phi(-v_{ij})} + \sum_{i=1}^{m}\sum_{j=1}^{r} v_{ij} = 0$$

and

$$-mr - \sum_{i=1}^{m}\sum_{j=1}^{r}(z_i^{(j)} + i - 1)\frac{v_{ij}\varphi(v_{ij})}{\Phi(v_{ij})} + \sum_{i=1}^{m}\sum_{j=1}^{r}(2m - z_i^{(j)} - i)\frac{v_{ij}\varphi(v_{ij})}{\Phi(-v_{ij})} + \sum_{i=1}^{m}\sum_{j=1}^{r} v_{ij}^2 = 0$$

where $\varphi(.)$ and $\Phi(.)$ are the pdf and the cdf of the standard normal distribution. The last two equations have no closed form solution; so numerical calculations are required to obtain the MLEs of $\mu$ and $\sigma$. A suitable and efficient software for solving such equations can be found in the R package *BB*, which is designed to solve systems of nonlinear equations.

**Large-sample distribution of the MLEs**

If $I_\mathcal{D}(\theta)$ denotes the Fisher information number about $\theta$ contained in $\mathcal{D}$, then according to Schervich (1995), $I_\mathcal{D}(\theta) = I_\mathcal{X}(\theta) + I_{\mathcal{Z}|\mathcal{X}}(\theta)$, where $\mathcal{X} = \left\{X_{i\pi(i)}^{(j)} | i = 1, 2, \ldots, m; j = 1, 2, \ldots, r\right\}$ and $\mathcal{Z} = \left\{Z_i^{(j)} | i = 1, 2, \ldots, m; j = 1, 2, \ldots, r\right\}$. Chen et al. (2004) give an expression for $I_\mathcal{X}(\theta)$ when the parent distribution is a location-scale, which is

$$I_\mathcal{X}(\theta) = mr I_{SRS}(\theta) + mr(r-1)\Delta(\theta),$$

where

$$\Delta(\theta) = E\left(\frac{\frac{\partial^2 F(X;\theta)}{\partial\theta \partial\theta^T}}{F(x;\theta)(1 - F(x;\theta))}\right),$$

where the expectation is taken over $X \sim F(x;\theta)$. For location-scale family,

$$\Delta(\theta) = \begin{pmatrix} \Delta_{11}(\theta) & \Delta_{12}(\theta) \\ \Delta_{12}(\theta) & \Delta_{22}(\theta) \end{pmatrix},$$

where

$$\Delta_{11}(\theta) = \frac{1}{\sigma^2} E\left(\frac{f(X,\theta)^2}{F(X,\theta)(1 - F(X,\theta))}\right), \Delta_{12}(\theta) = \frac{1}{\sigma^2} E\left(\frac{Xf(X,\theta)^2}{F(X,\theta)(1 - F(X,\theta))}\right)$$

and

$$\Delta_{22}(\boldsymbol{\theta}) = \frac{1}{\sigma^2} E\left(\frac{X^2 f(X,\boldsymbol{\theta})^2}{F(X,\boldsymbol{\theta})(1-F(X,\boldsymbol{\theta}))}\right).$$

For a symmetric distribution $\Delta_{12}(\boldsymbol{\theta}) = 0$. As a special case when the $F(.;.)$ is normal distribution, $\Delta(\boldsymbol{\theta})$ reduces to

$$\Delta(\boldsymbol{\theta}) = \begin{pmatrix} \frac{0.4805}{\sigma^2} & 0 \\ 0 & \frac{0.0675}{\sigma^2} \end{pmatrix}.$$

For one-parameters exponential distribution, i.e., $X \sim \exp(\sigma)$, $\Delta(\boldsymbol{\theta}) = 0.4041/\sigma^2$.

Now, we calculate $I_{\mathcal{Z}|\mathcal{Y}}(\boldsymbol{\theta})$. Since

$$\mathcal{Z}|\mathcal{X} \sim f_{\mathcal{Z}|\mathcal{X}}(\mathbf{z}|\mathcal{X};\boldsymbol{\theta}) = \prod_{i=1}^{m}\prod_{j=1}^{r} \binom{m}{z_i^{(j)}} F\left(\frac{x_{i\pi(i)}^{(j)}-\mu}{\sigma}\right)^{z_i^{(j)}} \left(1-F\left(\frac{x_{i\pi(i)}^{(j)}-\mu}{\sigma}\right)\right)^{m-z_i^{(j)}},$$

then

$$\frac{\partial}{\partial \mu}\log f_{\mathcal{Z}|\mathcal{X}}(\mathbf{z}|\mathcal{X};\boldsymbol{\theta}) = -\frac{1}{\sigma}\sum_{i=1}^{m}\sum_{j=1}^{r} z_i^{(j)}\frac{f(v_{ij})}{F(v_{ij})} + \frac{1}{\sigma}\sum_{i=1}^{m}\sum_{j=1}^{r}(m-z_i^{(j)})\frac{f(v_{ij})}{1-F(v_{ij})}$$

and

$$\frac{\partial^2}{\partial \mu^2}\log f_{\mathcal{Z}|\mathcal{X}}(\mathbf{z}|\mathcal{X};\boldsymbol{\theta}) = \frac{1}{\sigma^2}\sum_{i=1}^{m}\sum_{j=1}^{r} z_i^{(j)}\left(\frac{f'(v_{ij})}{F(v_{ij})} - \frac{f^2(v_{ij})}{F(v_{ij})^2}\right)$$

$$-\frac{1}{\sigma^2}\sum_{i=1}^{m}\sum_{j=1}^{r}(m-z_i^{(j)})\left(\frac{f'(v_{ij})}{1-F(v_{ij})} + \frac{f^2(v_{ij})}{\left(1-F(v_{ij})\right)^2}\right).$$

Taking the expectation over the pdf of $\mathcal{Z}|\mathcal{Y}$ yields:

$$E\left(\frac{\partial^2}{\partial \mu^2}\log f_{\mathcal{Z}|\mathcal{X}}(\mathbf{z}|\mathcal{X};\boldsymbol{\theta})\right) = -\frac{m}{\sigma^2}\sum_{i=1}^{m}\sum_{j=1}^{r} E\left(\frac{f^2(V_{ij})}{F(V_{ij})(1-F(V_{ij}))}\right).$$

So

$$I_{\mathcal{Z}|\mathcal{Y}}(\boldsymbol{\theta})_{11} = \frac{m}{\sigma^2}\sum_{i=1}^{m}\sum_{j=1}^{r} E\left(\frac{f^2(V_{ij})}{F(V_{ij})(1-F(V_{ij}))}\right) = \frac{mrA}{\sigma^2}$$

where $A = \sum_{i=1}^{m} E\left(\frac{f^2(W_i)}{F(W_i)(1-F(W_i))}\right)$ and $W_i \sim i\binom{m}{i} F(x)^{i-1}(1-F(x))^{m-i} f(x)$, for $i = 1, 2, \ldots, m$. Similarly,

$$\frac{\partial}{\partial \sigma} \log f_{Z|X}(z|X;\theta) = \frac{m}{\sigma} \sum_{i=1}^{m} \sum_{j=1}^{r} \frac{v_{ij} f(v_{ij})}{1-F(v_{ij})} - \frac{1}{\sigma} \sum_{i=1}^{m} \sum_{j=1}^{r} \frac{z_i^{(j)} v_{ij} f(v_{ij})}{F(v_{ij})(1-F(v_{ij}))}.$$

and

$$\frac{\partial^2}{\partial \sigma^2} \log f_{Z|X}(z|X;\theta) = -\frac{m}{\sigma^2} \sum_{i=1}^{m} \sum_{j=1}^{r} \left(\frac{v_{ij}^2 f'(v_{ij}) + v_{ij} f(v_{ij})}{1-F(v_{ij})} - \frac{v_{ij}^2 f^2(v_{ij})}{(1-F(v_{ij}))^2}\right) +$$

$$\frac{1}{\sigma^2} \sum_{i=1}^{m} \sum_{j=1}^{r} z_i^{(j)} \left(\frac{v_{ij}^2 f'(v_{ij}) - v_{ij} f(v_{ij})}{F(v_{ij})(1-F(v_{ij}))}\right.$$

$$\left. - \frac{v_{ij}^2 f^2(v_{ij}) - 2 v_{ij}^2 f^2(v_{ij}) F(v_{ij})}{F(v_{ij})^2 (1-F(v_{ij}))^2}\right).$$

Hence following similar procedure, we get

$$I_{Z|X}(\theta)_{22} = \frac{mrB}{\sigma^2} \text{ and } I_{Z|X}(\theta)_{12} = \frac{mrC}{\sigma^2},$$

where

$$B = \sum_{i=1}^{m} E\left(\frac{2W_i f(W_i) F(W_i)(1-F(W_i)) - 3W_i^2 f^2(W_i) F(W_i) + W_i^2 f^2(W_i)}{F(W_i)(1-F(W_i))^2}\right).$$

and $C = \sum_{i=1}^{m} E\left(\frac{W_i f^2(W_i)(3F(W_i)-1)}{F(W_i)(1-F(W_i))^2}\right)$. Applying Theorem 6.1. page 444 of Lehmann (1983), we find that $\sqrt{n}(\hat{\mu}_n - \mu) \xrightarrow{d} N(0, I_{\mathcal{D}}^{-1}(\theta)_{11})$ and $\sqrt{n}(\hat{\sigma}_n - \sigma) \xrightarrow{d} N(0, I_{\mathcal{D}}^{-1}(\theta)_{22})$, where $I_{\mathcal{D}}^{-1}(\theta)_{11}$ and $I_{\mathcal{D}}^{-1}(\theta)_{22}$ are the diagonal elements of $I_{\mathcal{D}}^{-1}(\theta)$.

### 3. An Illustrative Example

In order to select a sample via GRSS scheme, we propose the following approach. We select $2m$ observations say $X_1, X_2, \ldots, X_m, Y_1, Y_2, \ldots, Y_m$ from the population of interest and keep their actual measurements hidden to us. Then we plot $2m$ circles of radii $X_1, X_2, \ldots, X_m, Y_1, Y_2, \ldots, Y_m$ respectively. In Figure 1, the red circles represent $X_1, X_2, \ldots, X_m$ while the blue circles represent the $Y_1, Y_2, \ldots, Y_m$. Then via visual method we find the circle with radius $X_{(i)}$, which can be detected based on circles areas, and also, via the visual method we count how many of the blue circles have radii less than $X_{(i)}$, i.e., we measure $Z_i$. Then we repeat this procedure for $i = 1, 2, 3$. Then we get $(X_{1\pi(1)}, Z_1), (X_{2\pi(2)}, Z_2), \ldots, (X_{m\pi(m)}, Z_m)$ an GRSS sample of size $m$. Then the GRSS observed values are substituted in the likelihood equations $\frac{d\mathcal{L}^*(\boldsymbol{\theta};\mathcal{D})}{d\mu} = 0$ and $\frac{d\mathcal{L}^*(\boldsymbol{\theta};\mathcal{D})}{d\sigma} = 0$, with the $F(x; \boldsymbol{\theta}) = \Phi\left(\frac{x-\mu}{\sigma}\right)$. Solving for $\mu$ and $\sigma$ a numerical solution say $\hat{\mu}_{GRSS}$ and $\hat{\sigma}_{GRSS}$, for $\frac{d\mathcal{L}^*(\boldsymbol{\theta};\mathcal{D})}{d\mu} = 0$ and $\frac{d\mathcal{L}^*(\boldsymbol{\theta};\mathcal{D})}{d\sigma} = 0$ are obtained so that the hessian matrix at the solution is negative. The means square error (MSEs) for the estimators are obtained via the parametric bootstrap, i.e. $\widehat{\text{MSE}}(\hat{\theta}) = \frac{1}{N-1}\sum_{i=1}^{N}(\hat{\mu}_{GRSS_i} - \hat{\mu}_{GRSS})^2$, where $\hat{\mu}_{GRSS_i}$'s is a random from $N(\hat{\mu}_{GRSS}, \hat{\sigma}_{GRSS})$.

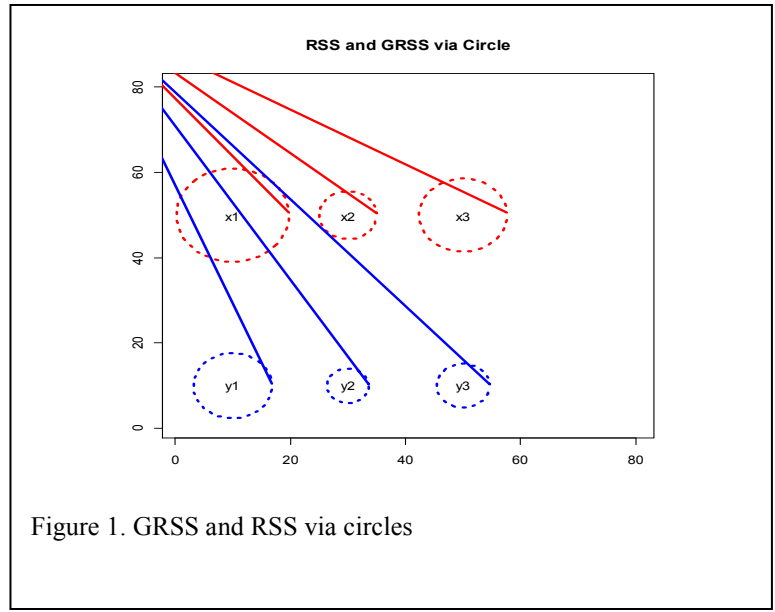

Figure 1. GRSS and RSS via circles

**Table 1.** GRSS obtained using circle method from four distributions.

| | $Normal(5,9)$ | | | | $Logistic(5,3)$ | | |
|---|---|---|---|---|---|---|---|
| Cycle | $(x_{1\pi(1)}, z_1)$ | $(x_{2\pi(2)}, z_2)$ | $(x_{3\pi(3)}, z_3)$ | Cycle | $(x_{1\pi(1)}, z_1)$ | $(x_{2\pi(2)}, z_2)$ | $(x_{3\pi(3)}, z_3)$ |
| 1 | (4.4474, 1) | (4.6752, 0) | (3.5313, 2) | 1 | (1.3716, 2) | (5.9190, 1) | (3.4550, 1) |
| 2 | (1.6178, 1) | (5.4656, 2) | (5.9896, 2) | 2 | (-0.164, 1) | (9.4621, 3) | (4.8668, 1) |
| 3 | (-1.0743, 0) | (8.460, 2) | (5.2522, 0) | 3 | (-2.022, 0) | (3.7473, 0) | (8.2979, 1) |
| 4 | (4.6148, 2) | (5.9695, 2) | (5.680, 3) | 4 | (9.4343, 3) | (6.3316, 2) | (10.9176, 3) |
| 5 | (-0.0245, 0) | (6.4033, 3) | (5.9212, 2) | 5 | (2.8089, 0) | (-0.688, 0) | (7.9176, 2) |
| | $Laplace(5,3)$ | | | | $Two-Prameter-Exp(5,3)$ | | |
| Cycle | $(x_{1\pi(1)}, z_1)$ | $(x_{2\pi(2)}, z_2)$ | $(x_{3\pi(3)}, z_3)$ | Cycle | $(x_{1\pi(1)}, z_1)$ | $(x_{2\pi(2)}, z_2)$ | $(x_{3\pi(3)}, z_3)$ |
| 1 | (1.6319, 0) | (8.7022, 3) | (9.9915, 3) | 1 | (6.2509, 2) | (7.1096, 2) | (9.3171, 2) |
| 2 | (6.5161, 2) | (5.0073, 0) | (15.141, 2) | 2 | (5.519, 1) | (8.2284, 3) | (6.250, 0) |
| 3 | (-6.0655, 0) | (5.2041, 2) | (10.064, 3) | 3 | (7.4602, 1) | (7.4863, 3) | (11.0521, 3) |
| 4 | (-3.7891, 1) | (8.8777, 2) | (5.9374, 2) | 4 | (5.2761, 0) | (5.3311, 0) | (19.0304, 3) |
| 5 | (0.1522, 1) | (3.3671, 2) | (9.0545, 3) | 5 | (6.2796, 1) | (5.6520, 1) | (14.0542, 3) |

**Table 2.** Biases and MSEs for MLE estimates for simulated data from four distributions

| Distribution | RSS | | GRSS | |
|---|---|---|---|---|
| | Location | Scale | Location | Scale |
| Normal(5,3) | 0.57216(0.2303) | 0.1239(0.1912) | 0.7424(0.1319) | 0.6171(0.1432) |
| Logisistic(5,3) | -0.166(0.9542) | -0.12699(0.3747) | -0.010(0.4893) | -0.092(0.2829) |
| Laplace(5,3) | -0.0546(0.3908) | -0.1357(0.4417) | -0.009(0.2418) | -0.1188(0.3342) |
| Two-param-Exp(5,3) | 0.1662(0.0565) | 0.3318(0.2373) | 0.1347(0.0427) | -0.1272(0.1519) |

## 4. A Simulation study

According to Chen et al. (2004), a RSS is expected to carry more information about the population of interest than the SRS of the same size. The MLE based on a RSS has been studied and compared extensively with its SRS counterpart (Chen et al., 2004). It has been is shown the MLE based on RSS is more efficient than its SRS counterpart. In RSS literature, it known that the statistical methods based on a RSS are tending to be superior to their SRS counterparts (Mahdizadeh and Zamanzade, 2017). Therefore, in this paper we will compare the MLE under the new proposed sampling with its counterpart using a RSS. So in this section we conduct a simulation study to compare the estimators of location and scale parameters under the GRSS with their RSS counterparts when the parent populations are $N(\mu,\sigma^2)$, $\text{Logist}(\mu,\sigma^2)$, $\text{Laplace}(\mu,\sigma^2)$ or Two-parameter Exponential $(\mu,\sigma^2)$. The estimators are compared via their biases (Bias) and mean square errors (MSE). The simulation results are presented in Tables 1 & 2 for $\mu = 5, \sigma = 3$ and different values of $n$ and $r$. In particular we used $m = 3$ with $r = 5, 10, 15, 20, 25$ and $m = 5$ with $r = 3, 6, 9, 12, 15$. The Bias and MSE for each estimators $\hat{\theta}$ are estimated as follows:

$$\widehat{\text{Bias}}(\hat{\theta}) = \frac{1}{N}\sum_{i=1}^{N}(\hat{\theta}_i - \theta) \text{ and } \widehat{\text{MSE}}(\hat{\theta}) = \frac{1}{N}\sum_{i=1}^{N}(\hat{\theta}_i - \theta)^2,$$

where $\hat{\theta}_1, \hat{\theta}_2, \ldots, \hat{\theta}_N$ are simulated values of the estimator $\hat{\theta}$.

Foss et al. (2013) defines the heavy-tailed distribution as follows. A pdf $f(x)$ is said to be heavy-tailed if it is not bounded by the exponential pdf, i.e. if $\lim_{x\to\infty} \sup f(x)e^{\gamma x} = \infty$ for all $\gamma > 0$. Also the pdf $f(x)$ is said to have a heavier tail than the pdf $g(y)$ if $\lim_{x\to\infty} \frac{f(x)}{f_Y(x)} = \infty$. Based on these two definitions, it can be seen that the Laplace and the Logistic distributions have heavier tails than the normal distribution. Also the Laplace distribution is heavier than the logistic distribution. The normal distribution is called a light-tailed distribution (Balakrishnan, 1992; Johnson et al., 1994). Therefore, these four distributions have been chosen, in our simulation study, since they have different types of tails. Furthermore, the kurtosis measure for these distributions varies from 3 to 9 as given in the Table 2.

**Table 3.** Kurtosis for the four distributions

| Distribution | Kurtosis |
|---|---|
| $N(0,1)$ | 3 |
| $\text{Logist}(\mu,\sigma^2)$ | 4.2 |
| $\text{Laplace}(\mu,\sigma^2)$ | 6 |
| Two-parameter Exponential $(\mu,\sigma^2)$ | 9 |

**Table 4.** Biases and Mean square errors for MLEs for different distributions, $m = 3$ and different $r = 5, 10, 15, 20, 25$.

| | | | | $N(5,9)$ | |
|---|---|---|---|---|---|
| | | | | *Location* | *Scale* |

|   |   |   | Bias |  | MSE |  | Bias |  | MSE |  |
|---|---|---|------|------|------|------|------|------|------|------|
| n | m | r | RSS | GRSS | RSS | GRSS | RSS | GRSS | RSS | GRSS |
| 15 | 3 | 5 | -0.0184 | -0.0118 | 0.3051 | 0.1801 | -0.0922 | -0.0670 | 0.2517 | 0.1895 |
| 30 | 3 | 10 | -0.0028 | 0.00710 | 0.1544 | 0.0923 | -0.0541 | -0.0488 | 0.1225 | 0.0925 |
| 45 | 3 | 15 | -0.0024 | -0.0058 | 0.1024 | 0.0591 | -0.0403 | -0.0276 | 0.0786 | 0.0590 |
| 60 | 3 | 20 | -0.0027 | -0.0037 | 0.0799 | 0.0449 | -0.0273 | -0.0219 | 0.0573 | 0.0451 |
| 75 | 3 | 25 | -0.0016 | -0.0008 | 0.0624 | 0.0358 | -0.0224 | -0.0143 | 0.0482 | 0.0370 |

|   |   |   | *Logistic*(5, 3) |  |  |  |  |  |  |  |
|---|---|---|---|---|---|---|---|---|---|---|
|   |   |   | Location |  |  |  | Scale |  |  |  |
|   |   |   | Bias |  | MSE |  | Bias |  | MSE |  |
| n | m | r | RSS | GRSS | RSS | GRSS | RSS | GRSS | RSS | GRSS |
| 15 | 3 | 5 | -0.0232 | -0.0269 | 0.9277 | 0.5206 | -0.0929 | -0.0755 | 0.3402 | 0.2353 |
| 30 | 3 | 10 | -0.0041 | -0.0075 | 0.4321 | 0.2409 | -0.0506 | -0.0396 | 0.1596 | 0.1165 |
| 45 | 3 | 15 | -0.0064 | 0.0061 | 0.2894 | 0.1665 | -0.0380 | -0.0285 | 0.1104 | 0.0829 |
| 60 | 3 | 20 | -0.0020 | 0.0032 | 0.2353 | 0.1360 | -0.0272 | -0.0194 | 0.0770 | 0.0584 |
| 75 | 3 | 25 | -0.0032 | -0.0019 | 0.1699 | 0.1019 | -0.0172 | -0.0148 | 0.0616 | 0.0469 |

|   |   |   | *Laplace*(5, 3) |  |  |  |  |  |  |  |
|---|---|---|---|---|---|---|---|---|---|---|
|   |   |   | Location |  |  |  | Scale |  |  |  |
|   |   |   | Bias |  | MSE |  | Bias |  | MSE |  |
| n | m | r | RSS | GRSS | RSS | GRSS | RSS | GRSS | RSS | GRSS |
| 15 | 3 | 5 | 0.0067 | 0.0056 | 0.4006 | 0.2313 | -0.1117 | -0.0945 | 0.4725 | 0.3469 |
| 30 | 3 | 10 | 0.0046 | 0.0049 | 0.2008 | 0.1121 | -0.0300 | -0.0274 | 0.2280 | 0.1705 |
| 45 | 3 | 15 | 0.0183 | 0.0162 | 0.1289 | 0.0762 | -0.0397 | -0.0331 | 0.1545 | 0.1161 |
| 60 | 3 | 20 | -0.0016 | -0.0054 | 0.0905 | 0.0530 | -0.0223 | -0.0184 | 0.1091 | 0.0798 |
| 75 | 3 | 25 | -0.0005 | 0.0038 | 0.0736 | 0.0417 | -0.0256 | -0.0222 | 0.0887 | 0.0658 |

|   |   |   | *Two-parameter-Exp*(5, 3) |  |  |  |  |  |  |  |
|---|---|---|---|---|---|---|---|---|---|---|
|   |   |   | Location |  |  |  | Scale |  |  |  |
|   |   |   | Bias |  | MSE |  | Bias |  | MSE |  |
| n | m | r | RSS | GRSS | RSS | GRSS | RSS | GRSS | RSS | GRSS |
| 15 | 3 | 5 | 0.1584 | 0.1219 | 0.0513 | 0.0360 | -0.3258 | -0.1202 | 0.2237 | 0.1482 |
| 30 | 3 | 10 | 0.0889 | 0.0757 | 0.0159 | 0.0126 | -0.0681 | 0.2264 | 0.0600 | 0.3627 |
| 45 | 3 | 15 | 0.0510 | 0.0527 | 0.0075 | 0.0062 | -0.0367 | -0.2487 | 0.2002 | 0.1364 |
| 60 | 3 | 20 | 0.0467 | 0.0416 | 0.0044 | 0.0038 | -0.3132 | -0.1339 | 0.1358 | 0.0458 |
| 75 | 3 | 25 | 0.0366 | 0.0334 | 0.0028 | 0.0024 | -0.2996 | -0.0406 | 0.1209 | 0.0325 |

**Table 5**. Biases and Mean square errors for MLEs for different distributions, $m = 5$ and different $r = 3, 6, 9, 12, 15$.

|   |   |   | $N(5, 9)$ |  |  |  |  |  |  |  |
|---|---|---|---|---|---|---|---|---|---|---|
|   |   |   | Location |  |  |  | Scale |  |  |  |
|   |   |   | Bias |  | MSE |  | Bias |  | MSE |  |
| n | m | r | RSS | GRSS | RSS | GRSS | RSS | GRSS | RSS | GRSS |
| 15 | 5 | 3 | 0.0017 | -0.0008 | 0.2097 | 0.1142 | -0.1057 | -0.0754 | 0.2075 | 0.1421 |
| 30 | 5 | 6 | 0.0012 | 0.0009 | 0.1017 | 0.0537 | -0.0554 | -0.0350 | 0.1004 | 0.0686 |

| n | m | r | Bias RSS | Bias GRSS | MSE RSS | MSE GRSS | Bias RSS | Bias GRSS | MSE RSS | MSE GRSS |
|---|---|---|---|---|---|---|---|---|---|---|
| 45 | 5 | 9 | 0.0006 | 0.0040 | 0.0674 | 0.0375 | -0.0330 | -0.0240 | 0.0679 | 0.0465 |
| 60 | 5 | 12 | 0.0003 | 0.0015 | 0.0511 | 0.0280 | -0.0276 | -0.0209 | 0.0495 | 0.0351 |
| 75 | 5 | 15 | 0.0016 | -0.0005 | 0.0411 | 0.0222 | -0.0194 | -0.0117 | 0.0388 | 0.0262 |

| | | | $Logistic(5,3)$ | | | | | | | |
|---|---|---|---|---|---|---|---|---|---|---|
| | | | Location | | | | Scale | | | |
| | | | Bias | | MSE | | Bias | | MSE | |
| n | m | r | RSS | GRSS | RSS | GRSS | RSS | GRSS | RSS | GRSS |
| 15 | 5 | 3 | 0.0115 | 0.0113 | 0.6104 | 0.3420 | -0.0796 | -0.0566 | 0.2563 | 0.1795 |
| 30 | 5 | 6 | -0.0018 | -0.0056 | 0.2994 | 0.1633 | -0.0455 | -0.0309 | 0.1314 | 0.0887 |
| 45 | 5 | 9 | -0.0051 | -0.0019 | 0.2083 | 0.1125 | -0.0202 | -0.0154 | 0.0904 | 0.0610 |
| 60 | 5 | 12 | 0.0022 | 0.0014 | 0.1524 | 0.0823 | -0.0266 | -0.0192 | 0.0634 | 0.0433 |
| 75 | 5 | 15 | -0.0059 | 0.0047 | 0.1250 | 0.0664 | -0.0174 | -0.0119 | 0.0528 | 0.0366 |

| | | | $Laplace(5,3)$ | | | | | | | |
|---|---|---|---|---|---|---|---|---|---|---|
| | | | Location | | | | Scale | | | |
| | | | Bias | | MSE | | Bias | | MSE | |
| n | m | r | RSS | GRSS | RSS | GRSS | RSS | GRSS | RSS | GRSS |
| 15 | 5 | 3 | -0.0201 | -0.0123 | 0.2720 | 0.1481 | -0.0553 | -0.0359 | 0.3721 | 0.2576 |
| 30 | 5 | 6 | -0.0036 | -0.0029 | 0.1303 | 0.0708 | -0.0387 | -0.0316 | 0.1850 | 0.1234 |
| 45 | 5 | 9 | -0.0030 | -0.0006 | 0.0843 | 0.0455 | -0.0230 | -0.0135 | 0.1305 | 0.0875 |
| 60 | 5 | 12 | 0.0046 | 0.0038 | 0.0628 | 0.0336 | -0.0159 | -0.0104 | 0.0989 | 0.0664 |
| 75 | 5 | 15 | -0.0027 | -0.0016 | 0.0488 | 0.0278 | -0.0137 | -0.0109 | 0.0714 | 0.0499 |

| | | | $Two\text{-}parameter\text{-}Exp(5,3)$ | | | | | | | |
|---|---|---|---|---|---|---|---|---|---|---|
| | | | Location | | | | Scale | | | |
| | | | Bias | | MSE | | Bias | | MSE | |
| n | m | r | RSS | GRSS | RSS | GRSS | RSS | GRSS | RSS | GRSS |
| 15 | 5 | 3 | 0.1407 | 0.0998 | 0.0421 | 0.0270 | -0.2707 | 0.0316 | 0.1946 | 0.1704 |
| 30 | 5 | 6 | 0.0778 | 0.0613 | 0.0127 | 0.0080 | -0.3609 | -0.2540 | 0.2692 | 0.1516 |
| 45 | 5 | 9 | 0.0560 | 0.0455 | 0.0066 | 0.0048 | -0.2820 | -0.0903 | 0.1355 | 0.0418 |
| 60 | 5 | 12 | 0.0426 | 0.0358 | 0.0037 | 0.0029 | -0.2639 | -0.0343 | 0.1206 | 0.0222 |
| 75 | 5 | 15 | 0.0356 | 0.0305 | 0.0026 | 0.0021 | -0.2395 | 0.1236 | 0.0972 | 0.0596 |

**Table 6**. Biases and Mean square errors for MLEs for different distributions, $m = 3$ and different $r = 5, 10, 15, 20, 25$.

| | | | $N(5, 0.04)$ | | | | | | | |
|---|---|---|---|---|---|---|---|---|---|---|
| | | | Location | | | | Scale | | | |
| | | | Bias | | MSE | | Bias | | MSE | |
| n | m | r | RSS | GRSS | RSS | GRSS | RSS | GRSS | RSS | GRSS |
| 15 | 3 | 5 | 0.0018 | 0.0007 | 0.0014 | 0.0008 | -0.0088 | -0.0065 | 0.0011 | 0.0008 |
| 30 | 3 | 10 | -0.0019 | -0.0012 | $7\times10^{-4}$ | $4\times10^{-4}$ | -0.0038 | -0.0030 | $5\times10^{-4}$ | $4\times10^{-4}$ |
| 45 | 3 | 15 | 0.0010 | 0.0004 | $4\times10^{-4}$ | $2\times10^{-4}$ | -0.0024 | -0.0019 | $4\times10^{-4}$ | $3\times10^{-4}$ |
| 60 | 3 | 20 | -0.0006 | -0.0002 | $3\times10^{-4}$ | $2\times10^{-4}$ | -0.0011 | -0.0007 | $3\times10^{-4}$ | $2\times10^{-4}$ |
| 75 | 3 | 25 | -0.0003 | -0.0003 | $3\times10^{-4}$ | $2\times10^{-4}$ | -0.0020 | -0.0011 | $2\times10^{-4}$ | $2\times10^{-4}$ |

| | | | $Logistic(5, 0.2)$ | | | | | | | |
|---|---|---|---|---|---|---|---|---|---|---|

|   |   |   | Location | | | | Scale | | | |
|---|---|---|---|---|---|---|---|---|---|---|
|   |   |   | Bias | | MSE | | Bias | | MSE | |
| n | m | r | RSS | GRSS | RSS | GRSS | RSS | GRSS | RSS | GRSS |
| 15 | 3 | 5 | 0.0005 | 0.0003 | 0.0042 | 0.0023 | -0.0060 | 0.0003 | 0.0015 | 0.0011 |
| 30 | 3 | 10 | 0.0003 | -0.0002 | 0.0020 | 0.0012 | -0.0030 | -0.0023 | 0.0008 | 0.0006 |
| 45 | 3 | 15 | 0.0002 | -0.0008 | 0.0013 | 0.0008 | -0.0027 | -0.0024 | 0.0005 | 0.0004 |
| 60 | 3 | 20 | -0.0009 | -0.0008 | $10\times10^{-4}$ | $6\times10^{-4}$ | -0.0016 | -0.0014 | $4\times10^{-4}$ | $3\times10^{-4}$ |
| 75 | 3 | 25 | -0.0006 | -0.0011 | $8\times10^{-4}$ | $5\times10^{-4}$ | -0.0015 | -0.0012 | $3\times10^{-4}$ | $2\times10^{-4}$ |
|   |   |   | $Laplace(5, 0.2)$ | | | | | | | |
|   |   |   | Location | | | | Scale | | | |
|   |   |   | Bias | | MSE | | Bias | | MSE | |
| n | m | r | RSS | GRSS | RSS | GRSS | RSS | GRSS | RSS | GRSS |
| 15 | 3 | 5 | -0.0014 | -0.0009 | 0.0019 | 0.0011 | -0.0069 | -0.0052 | 0.0022 | 0.0016 |
| 30 | 3 | 10 | 0.0003 | -0.0004 | 0.0009 | 0.0005 | -0.0027 | -0.0024 | 0.0011 | 0.0008 |
| 45 | 3 | 15 | -0.0005 | 0.0001 | $6\times10^{-4}$ | $3\times10^{-4}$ | -0.0020 | -0.0019 | $7\times10^{-4}$ | $5\times10^{-4}$ |
| 60 | 3 | 20 | -0.0002 | $6.7\times10^{-5}$ | $4\times10^{-4}$ | $2\times10^{-4}$ | -0.0017 | -0.0014 | $5\times10^{-4}$ | $4\times10^{-4}$ |
| 75 | 3 | 25 | 0.0002 | $-1.2\times10^{-4}$ | $3\times10^{-4}$ | $2\times10^{-4}$ | -0.0019 | -0.0019 | $4\times10^{-4}$ | $3\times10^{-4}$ |
|   |   |   | $Two\text{-}parameter\text{-}Exp(5, 0.2)$ | | | | | | | |
|   |   |   | Location | | | | Scale | | | |
|   |   |   | Bias | | MSE | | Bias | | MSE | |
| n | m | r | RSS | GRSS | RSS | GRSS | RSS | GRSS | RSS | GRSS |
| 15 | 3 | 5 | 0.0105 | 0.0084 | 0.0002 | 0.0002 | 0.0549 | 0.0024 | 0.0145 | 0.0044 |
| 30 | 3 | 10 | 0.0058 | 0.0049 | 0.0001 | 0.0001 | 0.1044 | 0.0090 | 0.0202 | 0.0040 |
| 45 | 3 | 15 | 0.0041 | 0.0035 | $3\times10^{-5}$ | $2.5\times10^{-5}$ | 0.0445 | 0.0104 | 0.0107 | 0.0046 |
| 60 | 3 | 20 | 0.0030 | 0.0027 | $1.9\times10^{-5}$ | $1.6\times10^{-5}$ | 0.0419 | -0.0028 | 0.0052 | 0.0029 |
| 75 | 3 | 25 | 0.0024 | 0.0022 | $1.3\times10^{-5}$ | $1.1\times10^{-5}$ | 0.0452 | 0.0034 | 0.0125 | 0.0035 |

**Table 7**. Biases and Mean square errors for MLEs for different distributions, $m = 5$ and different $r = 5, 10, 15, 20, 25$.

|   |   |   | $N(5, 0.04)$ | | | | | | | |
|---|---|---|---|---|---|---|---|---|---|---|
|   |   |   | Location | | | | Scale | | | |
|   |   |   | Bias | | MSE | | Bias | | MSE | |
| n | m | r | RSS | GRSS | RSS | GRSS | RSS | GRSS | RSS | GRSS |
| 15 | 5 | 3 | 0.0022 | 0.0002 | 0.0057 | 0.0032 | -0.0142 | -0.0098 | 0.0057 | 0.0040 |
| 30 | 5 | 6 | -0.0004 | 0.0001 | 0.0027 | 0.0016 | -0.0094 | -0.0058 | 0.0028 | 0.0020 |
| 45 | 5 | 9 | 0.0002 | -0.0004 | 0.0019 | 0.0011 | -0.0046 | -0.0041 | 0.0018 | 0.0013 |
| 60 | 5 | 12 | 0.0002 | 0.0002 | 0.0014 | 0.0008 | -0.0042 | -0.0031 | 0.0013 | 0.0010 |
| 75 | 5 | 15 | -0.0006 | -0.0003 | 0.0012 | 0.0007 | -0.0023 | -0.0013 | 0.0011 | 0.0007 |
|   |   |   | $Logistic(5, 0.2)$ | | | | | | | |
|   |   |   | Location | | | | Scale | | | |
|   |   |   | Bias | | MSE | | Bias | | MSE | |
| n | m | r | RSS | GRSS | RSS | GRSS | RSS | GRSS | RSS | GRSS |
| 15 | 5 | 3 | -0.0002 | -0.0002 | 0.0027 | 0.0014 | -0.0054 | -0.0041 | 0.0012 | 0.0008 |
| 30 | 5 | 6 | 0.0005 | -0.0002 | 0.0013 | 0.0007 | -0.0026 | -0.0020 | 0.0006 | 0.0004 |

| n | m | r | RSS | GRSS | RSS | GRSS | RSS | GRSS | RSS | GRSS |
|---|---|---|---|---|---|---|---|---|---|---|
| 45 | 5 | 9 | -0.0015 | -0.0004 | $9\times10^{-4}$ | $5\times10^{-4}$ | -0.0012 | -0.0008 | $4\times10^{-4}$ | $3\times10^{-4}$ |
| 60 | 5 | 12 | 0.0004 | 0.0005 | $6\times10^{-4}$ | $3\times10^{-4}$ | $-1\times10^{-3}$ | $-7\times10^{-4}$ | $3\times10^{-4}$ | $2\times10^{-4}$ |
| 75 | 5 | 15 | -0.0005 | -0.0001 | $5\times10^{-4}$ | $3\times10^{-4}$ | -0.0015 | -0.010 | $2\times10^{-4}$ | $1\times10^{-4}$ |

| | | | $Laplace(\mathbf{5},\mathbf{0.2})$ | | | | | | | |
|---|---|---|---|---|---|---|---|---|---|---|
| | | | Location | | | | Scale | | | |
| | | | Bias | | MSE | | Bias | | MSE | |
| n | m | r | RSS | GRSS | RSS | GRSS | RSS | GRSS | RSS | GRSS |
| 15 | 5 | 3 | 0.0011 | 0.0014 | 0.0012 | 0.0006 | -0.0037 | -0.0038 | 0.0018 | 0.0012 |
| 30 | 5 | 6 | -0.0004 | -0.0001 | $6\times10^{-4}$ | $3\times10^{-4}$ | -0.0029 | 0.0026 | $8\times10^{-4}$ | $6\times10^{-4}$ |
| 45 | 5 | 9 | -0.0008 | -0.0005 | $4\times10^{-4}$ | $2\times10^{-4}$ | -0.0015 | -0.0012 | $5\times10^{-4}$ | $4\times10^{-4}$ |
| 60 | 5 | 12 | $-1\times10^{-4}$ | $-1\times10^{-4}$ | $3\times10^{-4}$ | $2\times10^{-4}$ | $-7\times10^{-4}$ | $-7\times10^{-4}$ | $4\times10^{-4}$ | $3\times10^{-4}$ |
| 75 | 5 | 15 | -0.0002 | 0.0000 | $2\times10^{-4}$ | $1\times10^{-4}$ | -0.0015 | -0.0012 | $4\times10^{-4}$ | $2\times10^{-4}$ |

| | | | $Two\text{-}parameter\text{-}Exp(\mathbf{5},\mathbf{0.2})$ | | | | | | | |
|---|---|---|---|---|---|---|---|---|---|---|
| | | | Location | | | | Scale | | | |
| | | | Bias | | MSE | | Bias | | MSE | |
| n | m | r | RSS | GRSS | RSS | GRSS | RSS | GRSS | RSS | GRSS |
| 15 | 5 | 3 | 0.0088 | 0.0064 | 0.0002 | 0.0001 | 0.0042 | -0.0024 | 0.0044 | 0.0019 |
| 30 | 5 | 6 | 0.0051 | 0.0039 | 0.0001 | $3.0\times10^{-5}$ | 0.0143 | -0.0049 | 0.0037 | 0.0014 |
| 45 | 5 | 9 | 0.0036 | 0.0029 | $3\times10^{-5}$ | $2.0\times10^{-5}$ | 0.0238 | -0.0055 | 0.0043 | 0.0014 |
| 60 | 5 | 12 | 0.0028 | 0.0024 | $1.7\times10^{-5}$ | $1.3\times10^{-5}$ | 0.0135 | 0.0005 | 0.0027 | 0.0016 |
| 75 | 5 | 15 | 0.0024 | 0.0020 | $1.1\times10^{-5}$ | $0.87\times10^{-5}$ | 0.0064 | 0.0019 | 0.0025 | 0.0017 |

**Table 8**. Biases and Mean square errors for MLEs for different distributions, $m = 3$ and different $r = 5, 10, 15, 20, 25$.

| | | | $N(\mathbf{5},\mathbf{1})$ | | | | | | | |
|---|---|---|---|---|---|---|---|---|---|---|
| | | | Location | | | | Scale | | | |
| | | | Bias | | MSE | | Bias | | MSE | |
| n | m | r | RSS | GRSS | RSS | GRSS | RSS | GRSS | RSS | GRSS |
| 15 | 3 | 5 | 0.0015 | 0.0032 | 0.0347 | 0.0204 | -0.0366 | -0.0271 | 0.0266 | 0.0198 |
| 30 | 3 | 10 | -0.0009 | -0.0043 | 0.0165 | 0.0095 | -0.0240 | -0.0213 | 0.0127 | 0.0106 |
| 45 | 3 | 15 | 0.0040 | 0.0038 | 0.0105 | 0.0064 | -0.0137 | -0.0118 | 0.0093 | 0.0070 |
| 60 | 3 | 20 | -0.0015 | -0.0022 | 0.0088 | 0.0042 | -0.0133 | -0.0118 | 0.0068 | 0.0053 |
| 75 | 3 | 25 | -0.0029 | -0.0018 | 0.0065 | 0.0039 | -0.0085 | -0.0050 | 0.0048 | 0.0038 |

| | | | $Logistic(\mathbf{5},\mathbf{1})$ | | | | | | | |
|---|---|---|---|---|---|---|---|---|---|---|
| | | | Location | | | | Scale | | | |
| | | | Bias | | MSE | | Bias | | MSE | |
| n | m | r | RSS | GRSS | RSS | GRSS | RSS | GRSS | RSS | GRSS |
| 15 | 3 | 5 | 0.0110 | 0.0095 | 0.1027 | 0.0590 | -0.0364 | -0.0268 | 0.0338 | 0.0260 |
| 30 | 3 | 10 | 0.0020 | 0.0048 | 0.0489 | 0.02930 | -0.0169 | -0.01430 | 0.0182 | 0.0130 |
| 45 | 3 | 15 | -0.0011 | 0.0001 | 0.0334 | 0.0191 | -0.0119 | -0.0098 | 0.0124 | 0.0095 |
| 60 | 3 | 20 | 0.0017 | -0.0010 | 0.0257 | 0.0145 | -0.0092 | -0.0078 | 0.0093 | 0.0068 |
| 75 | 3 | 25 | -0.0009 | -0.0002 | 0.0202 | 0.0116 | -0.0052 | -0.0032 | 0.0071 | 0.0054 |

| | | | $Laplace(\mathbf{5},\mathbf{1})$ | | | | | | | |
|---|---|---|---|---|---|---|---|---|---|---|

|   |   |   | Location | | | | Scale | | | |
|---|---|---|---|---|---|---|---|---|---|---|
|   |   |   | Bias | | MSE | | Bias | | MSE | |
| n | m | r | RSS | GRSS | RSS | GRSS | RSS | GRSS | RSS | GRSS |
| 15 | 3 | 5  | -0.0029 | -0.0012 | 0.0459 | 0.0254 | -0.0354 | -0.0308 | 0.0512 | 0.0389 |
| 30 | 3 | 10 | -0.0024 | -0.0019 | 0.0222 | 0.0130 | -0.0136 | -0.010  | 0.0248 | 0.0186 |
| 45 | 3 | 15 | -0.0023 | -0.0016 | 0.0137 | 0.0079 | -0.0101 | -0.0053 | 0.0163 | 0.0126 |
| 60 | 3 | 20 | -0.0007 | -0.0001 | 0.0108 | 0.0063 | -0.0078 | -0.0077 | 0.0124 | 0.0091 |
| 75 | 3 | 25 | -0.0039 | -0.0011 | 0.0084 | 0.0049 | -0.0070 | -0.0062 | 0.0106 | 0.0078 |
|   |   |   | *Two-parameter-Exp*(5, 1) | | | | | | | |
|   |   |   | Location | | | | Scale | | | |
|   |   |   | Bias | | MSE | | Bias | | MSE | |
| n | m | r | RSS | GRSS | RSS | GRSS | RSS | GRSS | RSS | GRSS |
| 15 | 3 | 5  | 0.0529 | 0.0412 | 0.0056 | 0.0041 | 0.4055 | 0.0856 | 0.4451 | 0.1109 |
| 30 | 3 | 10 | 0.0301 | 0.0253 | 0.0018 | 0.0014 | 0.3521 | 0.0374 | 0.3984 | 0.0773 |
| 45 | 3 | 15 | 0.0198 | 0.0177 | 0.0008 | 0.0006 | 0.2150 | 0.0529 | 0.2597 | 0.0697 |
| 60 | 3 | 20 | 0.0154 | 0.0138 | 0.0005 | 0.0004 | 0.4416 | 0.0964 | 0.3848 | 0.0511 |
| 75 | 3 | 25 | 0.0125 | 0.0115 | 0.0003 | 0.0003 | 0.4654 | 0.0382 | 0.3635 | 0.0339 |

**Table 9**. Biases and Mean square errors for MLEs for different distributions, $m = 5$ and different $r = 5, 10, 15, 20, 25$.

|   |   |   | $N(5, 1)$ | | | | | | | |
|---|---|---|---|---|---|---|---|---|---|---|
|   |   |   | Location | | | | Scale | | | |
|   |   |   | Bias | | MSE | | Bias | | MSE | |
| n | m | r | RSS | GRSS | RSS | GRSS | RSS | GRSS | RSS | GRSS |
| 15 | 5 | 3  | -0.0057 | -0.0027 | 0.0217 | 0.0120 | -0.0318 | -0.0259 | 0.0223 | 0.0152 |
| 30 | 5 | 6  | 0.0005  | 0.0007  | 0.0114 | 0.0064 | -0.0169 | -0.0123 | 0.0114 | 0.0079 |
| 45 | 5 | 9  | 0.0001  | 0.0006  | 0.0074 | 0.0043 | -0.0079 | -0.0033 | 0.0071 | 0.0048 |
| 60 | 5 | 12 | 0.0019  | 0.0014  | 0.0055 | 0.0031 | -0.0091 | -0.0064 | 0.0055 | 0.0038 |
| 75 | 5 | 15 | 0.0005  | -0.0007 | 0.0045 | 0.0026 | -0.0062 | -0.0036 | 0.0042 | 0.0030 |
|   |   |   | *Logistic*(5, 1) | | | | | | | |
|   |   |   | Location | | | | Scale | | | |
|   |   |   | Bias | | MSE | | Bias | | MSE | |
| n | m | r | RSS | GRSS | RSS | GRSS | RSS | GRSS | RSS | GRSS |
| 15 | 5 | 3  | 0.0036  | 0.0079  | 0.0707 | 0.0312 | -0.0273 | -0.0207 | 0.0312 | 0.0209 |
| 30 | 5 | 6  | 0.0038  | 0.0005  | 0.0338 | 0.0145 | -0.0102 | -0.0065 | 0.0145 | 0.0101 |
| 45 | 5 | 9  | -0.0012 | -0.0018 | 0.0216 | 0.0098 | -0.0036 | -0.0022 | 0.0098 | 0.0065 |
| 60 | 5 | 12 | 0.0008  | -0.0010 | 0.0171 | 0.0074 | -0.0050 | -0.0046 | 0.0074 | 0.0047 |
| 75 | 5 | 15 | 0.0003  | -0.0011 | 0.0131 | 0.0058 | -0.0055 | -0.0023 | 0.0058 | 0.0040 |
|   |   |   | *Laplace*(5, 1) | | | | | | | |
|   |   |   | Location | | | | Scale | | | |
|   |   |   | Bias | | MSE | | Bias | | MSE | |
| n | m | r | RSS | GRSS | RSS | GRSS | RSS | GRSS | RSS | GRSS |
| 15 | 5 | 3 | 0.0033 | 0.0035 | 0.0297 | 0.0169 | -0.0228 | -0.0132 | 0.0422 | 0.0296 |
| 30 | 5 | 6 | 0.0037 | 0.0029 | 0.0153 | 0.0080 | -0.0165 | -0.0113 | 0.0204 | 0.0137 |

| n | m | r | \multicolumn{2}{c}{} | | | | | | |
|---|---|---|---|---|---|---|---|---|---|
| 45 | 5 | 9 | -0.0016 | -0.0007 | 0.0089 | 0.0049 | -0.0080 | -0.0080 | 0.0142 | 0.0094 |
| 60 | 5 | 12 | -0.0002 | 0.0005 | 0.0073 | 0.0039 | -0.0044 | -0.0055 | 0.0103 | 0.0071 |
| 75 | 5 | 15 | -0.0033 | -0.0010 | 0.0056 | 0.0030 | -0.0051 | -0.0035 | 0.0084 | 0.0058 |

| | | | \multicolumn{8}{c}{Two-parameter-Exp(5,1)} |
|---|---|---|---|---|---|---|---|---|---|---|
| | | | \multicolumn{4}{c}{Location} | \multicolumn{4}{c}{Scale} |
| | | | \multicolumn{2}{c}{Bias} | \multicolumn{2}{c}{MSE} | \multicolumn{2}{c}{Bias} | \multicolumn{2}{c}{MSE} |
| n | m | r | RSS | GRSS | RSS | GRSS | RSS | GRSS | RSS | GRSS |
| 15 | 5 | 3 | 0.0463 | 0.0323 | 0.0045 | 0.0028 | 0.1148 | -0.0209 | 0.1273 | 0.0369 |
| 30 | 5 | 6 | 0.0258 | 0.0197 | 0.0014 | 0.0010 | 0.0987 | 0.0047 | 0.0738 | 0.0236 |
| 45 | 5 | 9 | 0.0181 | 0.0148 | 0.0007 | 0.0005 | 0.1112 | 0.0279 | 0.1020 | 0.0240 |
| 60 | 5 | 12 | 0.0139 | 0.0115 | 0.0004 | 0.0003 | 0.0870 | 0.0446 | 0.0559 | 0.0160 |
| 75 | 5 | 15 | 0.0119 | 0.0101 | 0.0003 | 0.0002 | 0.0547 | 0.0081 | 0.0413 | 0.0137 |

## 5. Results and Conclusions

Based on the simulation results given in Table (3) to Table (9), it appears that the bias of the location estimator in GRSS is slightly higher than RSS in most cases. However, the MSE of the GRSS estimator is considerably smaller, which means that, overall, the GRSS based estimator better than the RSS based estimator. On the other hand, the estimator based on the GRSS for the scale parameter clearly better than the RSS based estimator with regards to both bias and MSE performance. This means that the GRSS procedure is successful in taking advantage of the additional information provided by the visual judgment at almost no cost.

Although we have applied the GRSS to location-scale family, the procedure can be easily generalized to the log location-scale family, i.e., the family of distributions described by:

$$f(x;\mu,\sigma) = \frac{1}{\sigma x} f\left(\frac{\log x - \mu}{\sigma}\right), \quad x > 0; \; -\infty < \mu < \infty, \sigma > 0.$$

This family includes the log-normal distribution, the log-logistic distribution and the smallest extreme value distributions. This generalization will extend the applicability of our results to the case of positive random variables which are of primary interest in reliability and survival analysis, see Lawless (2003) and Meeker and Escobar (1998).